\documentclass[%
 reprint,
superscriptaddress,
 amsmath,amssymb,
 aps,
prb,
]{revtex4-2}

\usepackage{color}
\usepackage{graphicx}% Include figure files
\usepackage{dcolumn}% Align table columns on decimal point
\usepackage{bm}% bold math
\usepackage{hyperref}% add hypertext capabilities
\hypersetup{colorlinks=true,linkcolor=blue,citecolor=blue,urlcolor=blue}
\usepackage[T1]{fontenc} 
\usepackage{newtxtext}   
\usepackage{newtxmath}   
\usepackage{upgreek}
\usepackage{mhchem}
\usepackage{tabularx}
\usepackage{textcomp}
\usepackage[section]{placeins}

\begin{document}

\preprint{APS/123-QED}

\title{Complex Temperature-dependent Thermal Conductivity in a Sawtooth Chain Magnet \ce{Fe2SiSe4}}  

\author{Kunya Yang}
\email{kunyang@sanxiau.edu.cn}
\affiliation{Department of Physics, Chongqing Three Gorges University, Chongqing 404100, China}
\affiliation{Low Temperature Physics Lab, College of Physics \& Center of Quantum Materials and Devices, Chongqing University, Chongqing 401331, China}

\author{Feihao Pan}
\affiliation{Laboratory for Neutron Scattering and Key Laboratory of Quantum State Construction and Manipulation (Ministry of Education), School of Physics, Renmin University of China, Beijing 100872, China}

\author{Liran Wang}
\affiliation{Institute for Quantum Materials and Technologies, Karlsruhe Institute of Technology, Kaiserstraße 12, 76131 Karlsruhe, Germany}

\author{Chenglin Shang}
\affiliation{Laboratory for Neutron Scattering and Key Laboratory of Quantum State Construction and Manipulation (Ministry of Education), School of Physics, Renmin University of China, Beijing 100872, China}

\author{Ying Zhu}
\affiliation{Low Temperature Physics Lab, College of Physics \& Center of Quantum
Materials and Devices, Chongqing University, Chongqing 401331, China}

\author{Xiancai Hu}
\affiliation{Low Temperature Physics Lab, College of Physics \& Center of Quantum
Materials and Devices, Chongqing University, Chongqing 401331, China}

\author{Sanjiang He}
\affiliation{Low Temperature Physics Lab, College of Physics \& Center of Quantum
Materials and Devices, Chongqing University, Chongqing 401331, China}

\author{Xinrun Mi}
\affiliation{Low Temperature Physics Lab, College of Physics \& Center of Quantum
Materials and Devices, Chongqing University, Chongqing 401331, China}
\affiliation{Chongqing Police College, Chongqing 401331, China}

\author{Long Zhang}
\affiliation{Low Temperature Physics Lab, College of Physics \& Center of Quantum
Materials and Devices, Chongqing University, Chongqing 401331, China}

\author{Aifeng Wang}
\affiliation{Low Temperature Physics Lab, College of Physics \& Center of Quantum
Materials and Devices, Chongqing University, Chongqing 401331, China}
\author{Yisheng Chai}
\affiliation{Low Temperature Physics Lab, College of Physics \& Center of Quantum
Materials and Devices, Chongqing University, Chongqing 401331, China}

\author{Frederic Hardy}
\affiliation{Institute for Quantum Materials and Technologies, Karlsruhe Institute of Technology, Kaiserstraße 12, 76131 Karlsruhe, Germany}

\author{Christoph Meingast}
\affiliation{Institute for Quantum Materials and Technologies, Karlsruhe Institute of Technology, Kaiserstraße 12, 76131 Karlsruhe, Germany}

\author{Peng Cheng}
\email{pcheng@ruc.edu.cn}
\affiliation{Laboratory for Neutron Scattering and Key Laboratory of Quantum State Construction and Manipulation (Ministry of Education), School of Physics, Renmin University of China, Beijing 100872, China}

\author{Mingquan He}
\email{mingquan.he@cqu.edu.cn}
\affiliation{Low Temperature Physics Lab, College of Physics \& Center of Quantum
Materials and Devices, Chongqing University, Chongqing 401331, China}

\date{\today}

\begin{abstract}
Geometrically frustrated magnets provide an ideal platform for exploring the interplay between lattice geometry and spin degrees of freedom. Here, we investigate the interactions between lattice and spin via thermal-transport measurements on the triangular sawtooth-lattice olivine magnet \ce{Fe2SiSe4}, which exhibits successive magnetic transitions at $T_1 = 110$ K (antiferromagnetic) and $T_2 = 50$ K (ferrimagnetic). Although phonons dominate the thermal conductivity, its temperature dependence displays a pronounced double-peak structure arising from spin–phonon coupling. In the intermediate temperature range between $T_1$ and $T_2$, resonant scattering of phonons by magnetic excitations around 5 meV produces a broad maximum around 60 K. Below $T_2$, the resonant spin–phonon scattering is strongly suppressed, leading to a rapid increase in thermal conductivity upon cooling and a pronounced low-temperature peak near 11 K, characteristic of heat transport governed by conventional phonon scattering mechanisms. Notably, this low-temperature peak is enhanced by a factor of $\sim5$ compared to the broad maximum at higher temperatures. These results demonstrate the strong sensitivity of thermal transport to spin–lattice interactions and highlight spin–phonon scattering as an effective mechanism for tailoring thermal conductivity in geometrically frustrated magnets.
\end{abstract}

\maketitle

\section{Introduction}
The complex interplay among lattice, spin, charge, and orbital degrees of freedom in geometrically frustrated magnets often gives rise to emergent quantum phases of matter, including unconventional superconductivity and quantum spin liquids \cite{wu2025,Zhang2025,yin2022,balents2010,zhao2019,syzranov2022}. Among such systems, the sawtooth chain—composed of a one-dimensional arrangement of corner-sharing triangles—has attracted considerable attention due to its close connection to flat-band physics, unconventional magnetism, and exotic magnetic excitations \cite{flat2010,FeOSeO2019}. As a fundamental building block of the well-known two-dimensional kagome lattice, the sawtooth chain hosts a rich phase diagram that is governed by the competition between intrachain and interchain exchange interactions \cite{Zhitomirsky2004,Richter2004}. The family of olivine chalcogenides with the general formula $A_2BX_4$ ($A$ = Mn, Fe, Co, Ni; $B$ = Si, Ge; $X$ = O, S, Se, Te) provides a prominent material realization of the sawtooth chain \cite{Geometric,ZnLS2006}. In these compounds, the geometrically frustrated coordination of the magnetic $A$ sites, combined with the sizable spin-orbit coupling associated with the transition-metal $A$ atoms, offers an ideal platform for investigating the interplay among multiple degrees of freedom \cite{Anomalous2005,chung2010magnetic,mandujano2023magnetic,chen2024}.

Of particular interest are the Fe-based olivines, owing to their strong spin-orbit and spin-phonon couplings \cite{FeGeCh2016,pan2023,schmidt1992magnon}. While \ce{Fe2SiS4} and \ce{Fe2SiO4} have been studied for decades \cite{baron1999neutron,aronson2007}, the single-crystal growth and detailed characterization of the structural and magnetic properties of \ce{Fe2SiSe4} were reported only recently by coauthors of this study \cite{pan2023}. As shown in Fig.~\ref{VSM}(a), \ce{Fe2SiSe4} crystallizes in the olivine-type structure with orthorhombic symmetry (space group $Pnma$, No. 62). Two crystallographically inequivalent Fe sites, Fe1 ($4a$ site) and Fe2 ($4c$ site), form sawtooth chains extending along the $b$ axis. The geometrical frustration inherent to this lattice, combined with complex intrachain and interchain exchange interactions, gives rise to successive magnetic transitions and a highly nontrivial magnetic ground state. Our previous powder neutron diffraction study revealed that \ce{Fe2SiSe4} undergoes an antiferromagnetic transition at $T_{1}$ = 110 K upon cooling \cite{pan2023}. The magnetic structure is characterized by a single-$\mathbf{q}$ propagation vector $\mathbf{q_1}=(0,0,0)$, with the magnetic moments on Fe2 (4.04 $\upmu_\mathrm{B}$/Fe) aligned along the $b$ axis, while the moments on Fe1 (1.45 $\upmu_\mathrm{B}$/Fe) are slightly canted toward the $a$ axis [see Fig.~\ref{VSM}(b)]. A second ferrimagnetic transition occurs at $T_{2}$ = 50 K, below which a double-$\mathbf{q}$ magnetic structure develops with $\mathbf{q_1}=(0,0,0)$ and $\mathbf{q_2}=(0,0.5,0)$ [see Fig.~\ref{VSM}(c)]. For the $\mathbf{q_1}$ component, the magnetic moments on Fe1 and Fe2 increase to 2.15 and 4.15 $\upmu_\mathrm{B}$/Fe, respectively. For the $\mathbf{q_2}$ component, Fe1 carries a magnetic moment of 2.46 $\upmu_\mathrm{B}$, whereas Fe2 exhibits two distinct moments of 2.75 and 0.03 $\upmu_\mathrm{B}$. A third magnetic transition at $T_{3}$ = 25 K is also observed in magnetization measurements [see Figs.~\ref{VSM}(d–f)], although its microscopic nature remains unclear. These complex magnetic transitions stand in sharp contrast to the behavior of \ce{Fe2SiS4} and \ce{Fe2SiO4}, which exhibit only a single-$\mathbf{q}$ antiferromagnetic transition \cite{Ehrenberg_1993,baron1999neutron}. 

In this article, we report comprehensive magnetization, specific-heat, thermal-expansion, and thermal-conductivity measurements on \ce{Fe2SiSe4} single crystals. In particular, the interplay between lattice and spin degrees of freedom gives rise to a complex temperature dependence of the thermal conductivity. Both above $T_1$ and below $T_2$, the thermal conductivity exhibits the characteristic behavior of phonon-dominated heat transport, with a pronounced peak near 11 K. In contrast, within the intermediate temperature range between $T_1$ and $T_2$, an additional broad maximum emerges in the thermal conductivity, which we attribute to resonant phonon scattering by magnetic excitations. The energy gap of these magnetic excitations is estimated to be $\Delta = 5.2$ meV ($\sim60$ K), which is in close proximity to the energy splitting induced by spin-orbit coupling of the Fe$^{2+}$ ions. Upon cooling through $T_2$, the concomitant emergence of the double-$\mathbf{q}$ magnetic order and sharp changes in lattice parameters likely lead to a mismatch between the magnetic excitation spectrum and the phonon modes, thereby restoring a conventional phonon-scattering mechanism. As a consequence, the peak thermal conductivity is enhanced by a factor of $\sim5$ due to the effective suppression of spin-phonon scattering. These results establish \ce{Fe2SiSe4} as a prominent platform for investigating the interplay among multiple degrees of freedom and demonstrate that thermal transport provides a highly sensitive probe for elucidating such interactions.

\section{Methods}
High-quality single crystals of \ce{Fe2SiSe4} were grown using the chemical vapor transport method \cite{pan2023}. All measurements were performed on the same crystal, with dimensions of $a \times b \times c = 1.1 \times 2.4 \times 0.7$ $\rm{mm}^3$. Magnetization, specific-heat, and electrical resistivity measurements were carried out using a Physical Property Measurement System (PPMS, Quantum Design Dynacool 9 T). Specific-heat measurements were performed using the relaxation method, while the electrical conductivity along the $b$ axis was measured using the standard four-probe technique.

Thermal-expansion measurements along different crystallographic axes were conducted using a home-made high-resolution capacitive dilatometer \cite{Meingast1990}. Steady-state thermal-transport measurements along the $b$ axis were carried out in the PPMS using a one-heater–two-thermometer configuration, with the thermometers calibrated in magnetic fields. The errors arising from heat losses in the thermal conductivity measurements are estimated to be below $\pm$ 4\% for temperatures below 150 K. Uncertainties associated with the determination of sample dimensions are approximately $\pm$ 10\%.

\section{Results and Discussion}

\begin{figure*}
    \centering
    \includegraphics[width=0.9\linewidth]{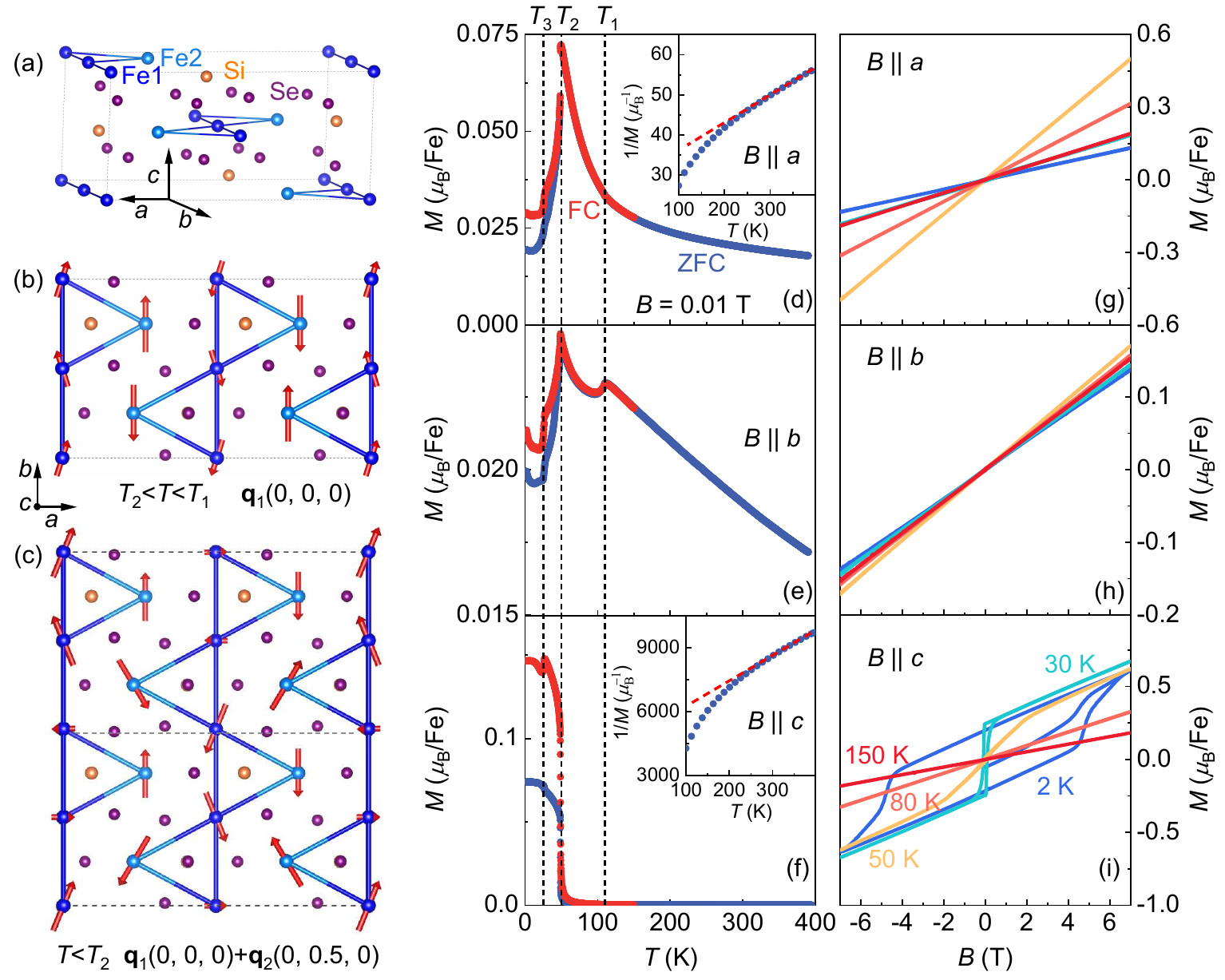}
   \caption{(a–c) Crystal and magnetic structures of \ce{Fe2SiSe4}. Two inequivalent Fe sites (Fe1 and Fe2) form a sawtooth-chain lattice. A single-$\mathbf{q}$ and a double-$\mathbf{q}$ magnetic structure emerge below $T_1=110$ K and $T_2=50$ K, respectively. (d–f) Temperature-dependent magnetization measured under zero-field-cooled (ZFC) and field-cooled (FC) protocols in a magnetic field of 0.01 T applied along different axes. Measurements were performed upon warming} (g–i) Isothermal magnetization measured with the magnetic field applied along the three crystallographic directions.
    \label{VSM}
\end{figure*}

Figures~\ref{VSM}(d–f) show the temperature-dependent magnetization $M(T)$ measured under zero-field-cooled (ZFC) and field-cooled (FC) conditions with an external magnetic field $B = 0.01$ T applied along the $a$, $b$, and $c$ axes, respectively. The quasi-one-dimensional sawtooth-chain structure gives rise to pronounced magnetic anisotropy even in the paramagnetic state.  The magnetization for $B \parallel c$ is two orders of magnitude smaller than those for $B \parallel a$ and $B \parallel b$. The much weaker susceptibility along the $c$ axis likely arises from an anisotropic $g$ factor induced by the low-symmetry crystal-field environment of the Fe$^{2+}$ ions. Similar $g$-factor-anisotropy-induced large susceptibility anisotropy has often been observed in frustrated magnets with low-symmetry crystal fields, such as $\alpha-$\ce{RuCl3}, \ce{Na2Co2TeO6}, and \ce{Na3Co2SbO6} \cite{Kubuta_aniso,Yao_aniso,Li_aniso}. In addition, anisotropic short-range spin correlations associated with anisotropic exchange interactions may persist in the paramagnetic phase and respond differently to fields applied along different crystallographic directions \cite{He_aniso}. As seen in the insets of Figs.~\ref{VSM}(d) and (f), the magnetization for $B \parallel a$ and $B \parallel c$ shows clear deviations from Curie--Weiss behavior below $\sim200$ K, likely associated with the presence of spin fluctuations. Notably, for $B \parallel b$, the magnetization increases nearly linearly upon cooling. Although such non-Pauli and non-Curie--Weiss linear temperature dependence of magnetization is rarely observed, it has been reported in iron-based superconductors and their nonsuperconducting parent compounds \cite{WangBa122,YanSr122,Klingeler1111,Li2018122,Hardy201911}. In those systems, however, the magnetization increases linearly with increasing temperature, a behavior that has been attributed to the suppression of antiferromagnetic fluctuations at elevated temperatures \cite{Zhang_2009linear}. In contrast, the linear decrease of magnetization with increasing temperature observed in \ce{Fe2SiSe4} for $B \parallel b$ suggests that antiferromagnetic fluctuations alone cannot account for this unusual behavior. An additional chain-direction-dependent contribution associated with anisotropic exchange interactions and/or spin-orbit coupling may also be relevant. Further experimental and theoretical studies are required to elucidate the microscopic origin of this unusual temperature dependence.

The antiferromagnetic transition at $T_1 = 110$ K is most clearly manifested for $B \parallel b$. Our previous neutron-diffraction results indicate that the ordered antiferromagnetic moments are predominantly aligned along the $b$ axis [see Fig.~\ref{VSM}(b)] \cite{pan2023}. A field applied along the $b$ axis couples most sensitively to the development of the antiferromagnetic order parameter, making the anomaly at $T_1 = 110$ K most pronounced. For fields applied along $a$ and $c$, the response is transverse to the dominant ordered moment and is therefore much less sensitive to the transition. Such direction-selective visibility of an antiferromagnetic transition in magnetization is commonly observed in anisotropic antiferromagnets, such as $\alpha-$\ce{Gd2S3} and \ce{MnBi2Se4} \cite{Clark_afm,EBISU_afm}. For $B \parallel a$ and $B \parallel b$, the temperature-dependent magnetization exhibits a peak at $T_2 = 50$ K, below which a clear separation between the ZFC and FC curves emerges. In contrast, for $B \parallel c$, the magnetization displays a ferromagnetic-like transition at $T_2$, which has been identified as ferrimagnetic in nature \cite{pan2023}.  Our previous neutron-diffraction study found no $c$-axis magnetic moment at the Fe sites in the single-$\mathbf{q}$ phase, whereas Fe2 develops a $c$-axis moment of approximately 1.14 $\upmu_\mathrm{B}$ in the double-$\mathbf{q}$ phase \cite{pan2023}. The isothermal magnetization curves shown in Figs.~\ref{VSM}(g--i) reveal that the magnetization increases linearly with magnetic field up to 7 T at all measured temperatures for both $B \parallel a$ and $B \parallel b$. In contrast, for $B \parallel c$, a clear hysteresis loop develops below $T_2$, with an enhanced coercive field at lower temperatures. This further supports the presence of a ferrimagnetic component along the $c$ axis.   Signatures of a third transition at $T_3 = 25$ K are visible in $M(T)$ for all three field directions, although its microscopic origin remains unresolved.

\begin{figure*}
    \centering
    \includegraphics[width=0.7\linewidth]{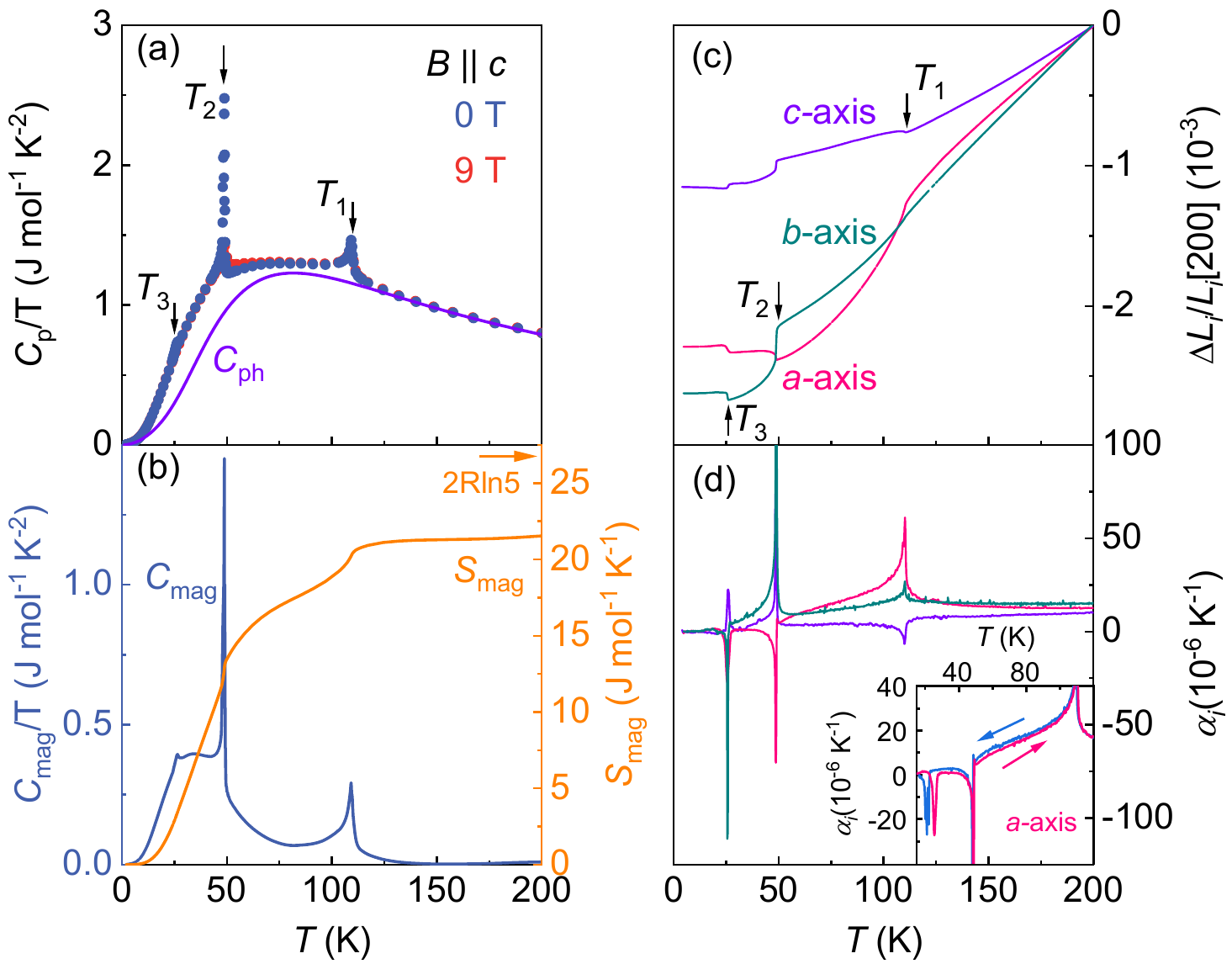}
    \caption{(a) Temperature dependence of the specific heat $C_\mathrm{p}$ (solid circles) measured upon warming in 0 and 9 T ($B \parallel c$) . The purple solid line represents the fitted phonon background using the Debye–Einstein model described in Eq.~\ref{eq:Cp_Fitting}. (b) Magnetic specific heat $C_\mathrm{mag}$ (blue line) obtained by subtracting the phonon contribution $C_\mathrm{ph}$ from the total $C_\mathrm{p}$ in zero magnetic field. The corresponding magnetic entropy $S_\mathrm{mag}$ is also shown (orange line). (c,d) Linear thermal expansion ($\Delta L_i/L_i$, $i=a,b,c$) and the corresponding thermal-expansion coefficient [$\alpha_i = (1/L_i)dL_i/dT$] measured along the three principal crystallographic axes upon warming. The inset in (d) compares the thermal-expansion coefficient along the $a$ axis measured during warming and cooling cycles.}
    \label{Cp}
\end{figure*}

Figure~\ref{Cp} presents the specific-heat and thermal-expansion data of \ce{Fe2SiSe4}. In zero magnetic field, sharp peak anomalies are observed in the specific heat at $T_1$ and $T_2$, whereas only a weak shoulder feature appears at $T_3$. The application of magnetic fields up to 9 T ($B \parallel c$) slightly suppresses the magnitude of the peak at $T_2$, while having a negligible effect on the transition temperatures $T_1$ and $T_2$. A magnetic field of 9 T along the $c$ axis is evidently insufficient to induce substantial changes in the magnetic structure, consistent with the relatively high transition temperatures and large exchange interactions (up to $\sim170$ K) reported previously \cite{pan2023}. To extract the magnetic contribution to the specific heat, $C_\mathrm{mag}$, we approximate the phonon contribution $C_\mathrm{ph}$ using the Debye–Einstein model:
\begin{equation}
\begin{split}
    C_\mathrm{ph} &= 9\alpha_\mathrm{D}NR\left(\frac{T}{\Theta_\mathrm{D}}\right)^3
                   \int_{0}^{\Theta_\mathrm{D}/T}\frac{x^{4}e^{x}}{(e^{x}-1)^{2}}dx \\
                  &+ 3(1-\alpha_\mathrm{D})NR\left(\frac{\Theta_\mathrm{E}}{T}\right)^{2}
                   \frac{e^{\Theta_\mathrm{E}/T}}{\left(e^{\Theta_\mathrm{E}/T}-1\right)^{2}},
\end{split}
\label{eq:Cp_Fitting}
\end{equation}
where $x = \hbar \omega / k_\mathrm{B} T$, $\omega$ is the phonon frequency, $k_\mathrm{B}$ is the Boltzmann constant, $\Theta_\mathrm{D}$ is the Debye temperature, $\Theta_\mathrm{E}$ is the Einstein temperature, $\alpha_\mathrm{D}$ is a fitting constant, $R$ is the ideal gas constant, and $N$ denotes the number of atoms per unit cell ($N = 7$ for \ce{Fe2SiSe4}). The specific heat in the paramagnetic state is well described by the Debye–Einstein phonon contribution with $\Theta_\mathrm{D} = 280$ K, $\Theta_\mathrm{E} = 235$ K, and $\alpha_\mathrm{D} = 0.75$, as shown by the purple solid line in Fig.~\ref{Cp}(a). The electronic contribution is negligible due to the relatively large band gap of $\sim0.66$ eV. The magnetic specific heat, plotted as the blue line in Fig.~\ref{Cp}(b), is then obtained by subtracting the phonon contribution from the total specific heat $C_\mathrm{p}$ according to $C_\mathrm{mag} = C_\mathrm{p} - C_\mathrm{ph}$. The magnetic specific heat extends well above $T_1$ and persists up to approximately 130 K, likely reflecting the thermal population of partially occupied spin-orbit manifolds and short-range spin correlations \cite{aronson2007}. Consistent with this picture, the magnetic entropy $S_\mathrm{mag}$ does not recover the ideal value of $2R\ln5$ expected for fully occupied states with $2S + 1 = 5$, even at temperatures well above $T_1$. 

Notably, the shoulder-like anomaly near $T_3$ becomes more pronounced in $C_\mathrm{mag}$. A similar feature has been reported in \ce{Fe2SiO4} around 20 K and has been attributed either to a spin-canting transition \cite{SANTORO1966655,Ehrenberg_1993,Lottermoser1995,Lottermoser1996} or to Schottky-type contributions arising from the lowest-lying magnetic excitations within the spin-orbit manifold of the Fe1 site \cite{aronson2007,HAFNER1990203}. Our previous neutron-diffraction study did not reveal any detectable change in the magnetic propagation vectors or in the ordered magnetic structure below $T_3$ \cite{pan2023}. Therefore, although a very small canting change below the resolution of the neutron data cannot be fully excluded, the anomaly at $T_3$ is unlikely to originate from a canting transition analogous to that discussed for \ce{Fe2SiO4}. Moreover, a sharp and hysteretic anomaly with a thermal width of approximately 4.5 K is observed around $T_3$ in the thermal-expansion coefficient of \ce{Fe2SiSe4} [see the inset of Fig.~\ref{Cp}(d)]. This behavior indicates that the transition at $T_3$ is of first order and is therefore unlikely to originate from the thermal population of low-lying excited states. The sharp hysteretic anomaly in thermal expansion, together with the relatively weak feature in the specific heat, suggests a weak first-order structural or magnetoelastic transition within the preexisting double-$\mathbf{q}$ magnetic state. The weak specific-heat signature indicates that the associated entropy change is small, as expected for a subtle lattice distortion, local magnetoelastic rearrangement, or domain/strain reconfiguration that preserves the average magnetic structure resolved by neutron diffraction. 

The linear thermal expansion ($\Delta L_i/L_i$, $i = a,b, c$) and the corresponding linear thermal-expansion coefficient [$\alpha_i = (1/L_i) dL_i/dT$] measured along the three crystallographic axes are shown in Figs.~\ref{Cp}(c,d). For all directions, a kink rather than a discontinuous jump is observed in $\Delta L_i/L_i$ at $T_1$. Consistently, the thermal-expansion coefficient $\alpha_i$ exhibits a pronounced $\lambda$-shaped anomaly at $T_1$. These features indicate that the transition at $T_1$ is of second order. In contrast, at $T_2$ and $T_3$, $\Delta L_i/L_i$ displays clear discontinuous jumps accompanied by sharp peaks in $\alpha_i$. As shown in the inset of Fig.~\ref{Cp}(d), no discernible hysteresis between cooling and warming is observed in the vicinity of $T_2$, suggesting that this transition is weakly first order. Notably, the substantial changes in lattice parameters, particularly along the $b$ axis, are expected to have a significant impact on thermal transport, as discussed below.  By comparison, the transition near $T_3$ exhibits pronounced hysteresis spanning approximately $\sim4.5$ K between measurements performed upon heating and cooling. This behavior unambiguously demonstrates that the transition at $T_3$ is of first order. Note that the magnetization and specific-heat measurements shown in Figs.~\ref{VSM} and \ref{Cp}, respectively, were performed upon warming; both show a transition at $T_3\sim25$ K, consistent with the warming branch of the thermal-expansion data.   

On the basis of the specific-heat and linear thermal-expansion data, we estimate the pressure dependence of the transition temperatures using the Clausius–Clapeyron relation:
\begin{equation}
    \frac{dT_{n}}{dp_i}= V_m\frac{\Delta L_i/L_i}{\Delta S},
    \label{Clausius}
\end{equation}
for a first-order phase transition, and the Ehrenfest relation:
\begin{equation}
    \frac{dT_{n}}{dp_i}= V_m\frac{\Delta \alpha_i}{\Delta C_\mathrm{p}/T_n},
    \label{Ehrenfest}
\end{equation}
for a second-order phase transition. Here, $dT_n/dp_i$ denotes the uniaxial ($i = a, b, c$) pressure dependence of the transition temperature $T_n$ ($n = 1, 2$), and $V_m = 36.2\ \mathrm{cm}^3\ \mathrm{mol}^{-1}$ is the molar volume of \ce{Fe2SiSe4}. The quantities $\Delta L_i$, $\Delta S$, $\Delta \alpha_i$, and $\Delta C_\mathrm{p}$ represent the discontinuities in the sample length, entropy, thermal-expansion coefficient, and specific heat at the corresponding transitions, respectively. The estimated uniaxial pressure derivatives, together with the hydrostatic pressure dependence [$dT_n/dp_h = dT_n/dp_a + dT_n/dp_b + dT_n/dp_c$], of the transition temperatures $T_n$ ($n = 1, 2$) are summarized in Table~\ref{Pressure}. The transition temperatures $T_1$ and $T_2$ are most sensitive to uniaxial pressure applied along the $a$ and $b$ axes, respectively, with both exhibiting large pressure derivatives of approximately 7 K GPa$^{-1}$.

\begin{table*}
\centering
\caption{Pressure derivatives of the magnetic transition temperatures $T_n$ ($n=1,2$) in \ce{Fe2SiSe4} estimated from thermal-expansion and specific-heat measurements.}
\label{Pressure}
\renewcommand*{\arraystretch}{1.5}% 
\begin{tabularx}{1\textwidth}{
>{\centering\arraybackslash}X
>{\centering\arraybackslash}X
>{\centering\arraybackslash}X 
>{\centering\arraybackslash}X
>{\centering\arraybackslash}X}
\hline
\hline
   & $a$ & $b$ & $c$ & $h$\\
\hline
$dT_1/dp_i$ K Gpa$^{-1}$ & 7.2(2) & 2.2(4) & -2.1(1) & 7.1(5)\\
$dT_2/dp_i$ K Gpa$^{-1}$ & -1.6(4) & 7.8(1) & 1.1(1) & 7.3(2)\\
\hline
\hline
\end{tabularx}
\end{table*}

\begin{figure*}
    \centering
    \includegraphics[width=0.7\linewidth]{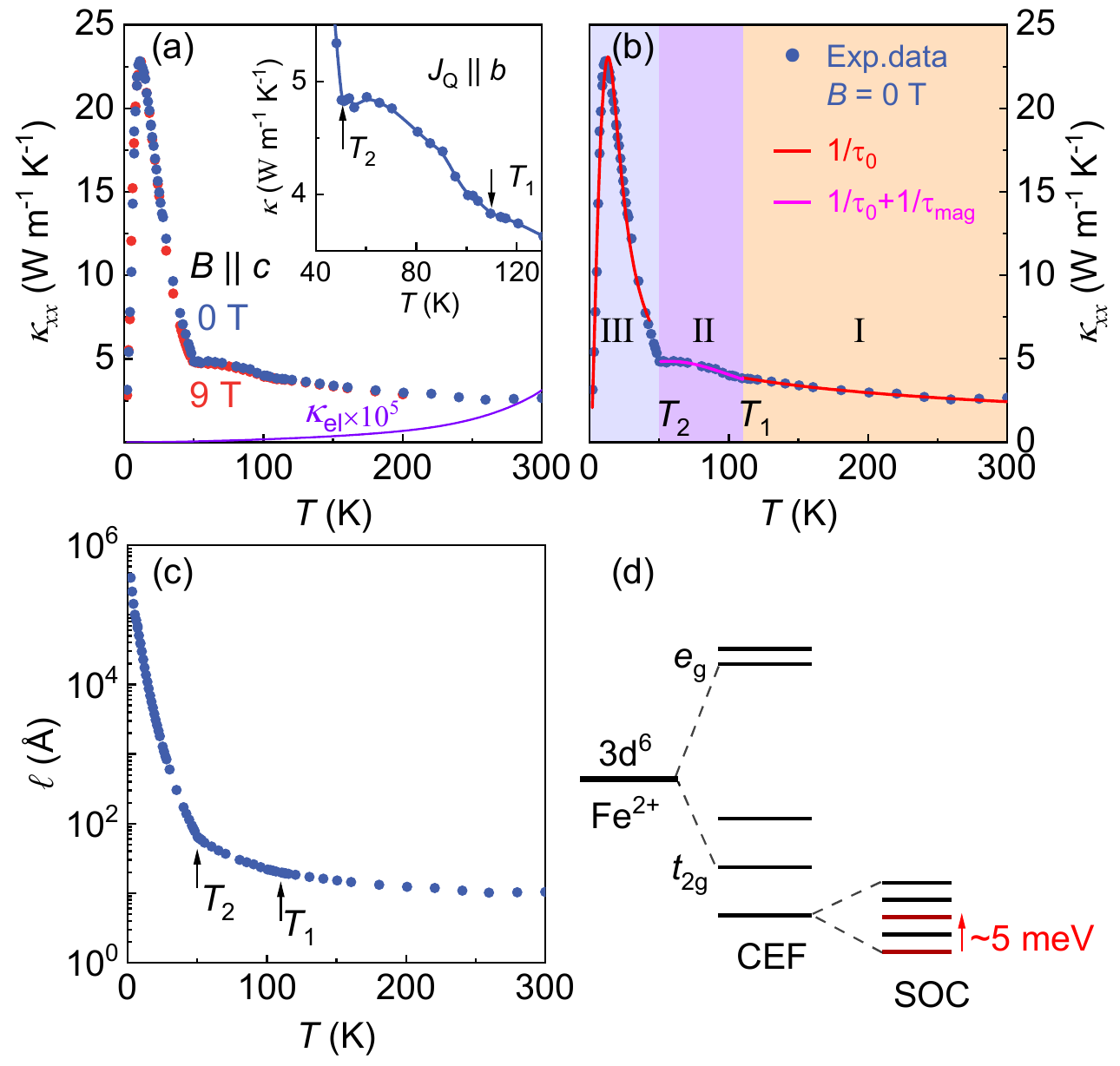}
    \caption{(a) Temperature dependence of the in-plane thermal conductivity $\kappa_{xx}$ of \ce{Fe2SiSe4}, measured with the heat current applied along the chain direction ($\mathbf{J}_\mathrm{Q} \parallel b$). A pronounced double-peak structure is observed in $\kappa_{xx}(T)$. Inset: enlarged view of the broad maximum around 60 K. The purple line denotes the electronic thermal conductivity $\kappa_\mathrm{el}$, multiplied by a factor of $10^{5}$ for visibility. (b) Fits to the temperature-dependent thermal conductivity using the Debye–Callaway model. In regions I and III, the data are well described by considering only the conventional phonon scattering rate $1/\tau_0$. In region II, an additional resonant spin–phonon scattering term $1/\tau_\mathrm{mag}$ is included to account for the broad maximum around 60 K. (c) Temperature dependence of the estimated phonon mean free path $\ell$. (d) Schematic illustration of the energy-level splitting of Fe$^{2+}$ ions induced by the crystal electric field (CEF) and spin–orbit coupling (SOC), adapted from Ref.~\cite{aronson2007}. Magnetic excitations at $\sim5$ meV resonantly scatter phonons, giving rise to the broad maximum in $\kappa_{xx}$ near 60 K.} 
    \label{k-T}
\end{figure*}

Figure~\ref{k-T} presents the complex temperature dependence of the longitudinal thermal conductivity $\kappa_{xx}(T)$ of \ce{Fe2SiSe4}, measured with the heat current applied along the $b$ axis. Reliable thermal-transport measurements along the $a$ and $c$ directions are challenging due to the limited sample dimensions along these axes. The thermal conductivity of \ce{Fe2SiSe4} is dominated by phononic heat transport ($\kappa_\mathrm{ph}$), while the electronic and magnetic contributions are negligible. The electronic thermal conductivity $\kappa_{\mathrm{el}}$, estimated using the Wiedemann–Franz law $\kappa_{\mathrm{el}} = L_0 T \sigma_{xx}$, does not exceed $5\times10^{-5}$ $\mathrm{W}\ \mathrm{m}^{-1}\ \mathrm{K}^{-1}$ below room temperature, as indicated by the purple line in Fig.~\ref{k-T}(a). Here, the electrical conductivity is obtained from the measured electrical resistivity via $\sigma_{xx} = 1/\rho_{xx}$ (data not shown). The magnetic contribution to the thermal conductivity, $\kappa_{\mathrm{mag}}$, is expected to scale with the magnetic specific heat according to $\kappa_{\mathrm{mag}} \propto C_{\mathrm{mag}} v_{\mathrm{mag}} \ell_{\mathrm{mag}}$, where $v_{\mathrm{mag}}$ and $\ell_{\mathrm{mag}}$ denote the group velocity and mean free path of magnetic heat carriers, respectively. Between $T_1$ and $T_2$, $C_{\mathrm{mag}}$ exhibits a pronounced dip, whereas $\kappa_{xx}$ shows a broad maximum. Furthermore, the shoulder-like anomaly in $C_{\mathrm{mag}}$ near $T_3$ has little influence on $\kappa_{xx}$. In addition, the application of a magnetic field of 9 T ($B \parallel c$) produces only a weak effect on the thermal conductivity. These observations indicate that $\kappa_{\mathrm{mag}}$ makes only a minor contribution to the total thermal conductivity, even within the magnetically ordered states.

In contrast to conventional phonon-dominated heat transport, the most striking feature of the thermal conductivity of \ce{Fe2SiSe4} is the appearance of a double-peak structure in its temperature dependence. As shown in Fig.~\ref{k-T}(a), a broad maximum develops around 60 K, followed by a second, more pronounced peak near 11 K. Similar double-peak temperature dependences of the thermal conductivity have been reported in several low-dimensional magnetic systems, including the quasi-two-dimensional antiferromagnets \ce{K2V3O8} and \ce{Nd2CuO4} \cite{Sales2002,JinNdCuO}, the spin-Peierls compound \ce{CuGeO3} \cite{CuGeO1998}, the Shastry–Sutherland compound \ce{SrCu2(BO3)2} \cite{Hofmann2001SrCu}, spin-chain compounds $AB_2\mathrm{O}_6$ ($A=$ Ni, Co; $B=$ Sb, Ta) \cite{PrasaiABO}, the Kitaev material $\alpha$-\ce{RuCl3} \cite{Hentrich2018}, and the layered van der Waals magnet \ce{Cr2Si2Te6} \cite{yang2023spin}. It is commonly believed that strong spin–phonon interactions are responsible for such double-peak temperature dependences of the thermal conductivity in these systems \cite{Sales2002,JinNdCuO,CuGeO1998,Hofmann2001SrCu,PrasaiABO,Hentrich2018,yang2023spin}. Similar spin–phonon coupling effects are therefore likely at play in \ce{Fe2SiSe4}.

We now analyze the distinct thermal-transport behaviors in different magnetic phases. For simplicity, we denote the paramagnetic phase as region I ($T>T_1$), the single-$\mathbf{q}$ phase as region II ($T_1>T>T_2$), and the double-$\mathbf{q}$ phase as region III ($T<T_2$) [see Fig.~\ref{k-T}(b)]. Because no discernible changes are observed in either the magnetic structure or the thermal conductivity at $T_3$, the phase below $T_3$ is also assigned to region III.  Since the thermal conductivity is dominated by phonons, its complex temperature dependence is governed by different phonon-scattering mechanisms in distinct temperature regimes. Within the kinetic phonon-gas picture, phonon heat transport can be well described by the Debye–Callaway model:
\cite{Callaway1959}:
\begin{equation}
    \kappa_\mathrm{ph}(T) = \frac{k_B^4}{2\pi^2 v_s\hbar^3} T^3 \int_0^{\Theta_\mathrm{D}/T} \frac{x^4 e^x}{(e^x-1)^2} \tau(\omega,T)dx,
    \label{Debye-calloway model}
\end{equation}
where $x = \hbar \omega / k_B T$, $v_s$ is the average sound velocity, and $\tau(\omega,T)$ denotes the phonon relaxation time.

\textit{Region I:} In region I, \ce{Fe2SiSe4} is in the paramagnetic phase, and only conventional phonon-scattering processes—including Umklapp scattering ($\tau_\mathrm{U}$), point-defect scattering ($\tau_\mathrm{pd}$), and boundary scattering ($\tau_\mathrm{b}$)—are expected to contribute. The total conventional scattering rate ($\tau_0^{-1}$) can be modeled using the Matthiessen’s rule by assuming independent scattering channels:
\begin{equation}
    \tau_0^{-1}=\tau^{-1}_\mathrm{b}+\tau^{-1}_\mathrm{pd}+\tau^{-1}_\mathrm{U}=v_sd^{-1}+P_d\omega^4+U_1\omega^2Te^{-\Theta_\mathrm{D}/U_2T},
\end{equation}
where $d$ is the effective boundary-scattering length scale, and $P_d$, $U_1$, and $U_2$ are numerical constants. As shown by the red solid line in Fig.~\ref{k-T}(b), the measured thermal conductivity in region I is well reproduced by the Debye–Callaway model when only the conventional scattering rate $\tau_0^{-1}$ is considered. The corresponding fitting parameters are summarized in Table~\ref{Debye-Callaway}.

\textit{Region II:}  Upon entering the single-$\mathbf{q}$ antiferromagnetic region II, the temperature dependence of the thermal conductivity deviates from the trend observed in region I. The emergence of the broad maximum around 60 K is attributed to resonant scattering of phonons by magnetic excitations \cite{Sales2002,JinNdCuO,CuGeO1998,Hofmann2001SrCu,PrasaiABO,Hentrich2018,yang2023spin}:
\begin{equation}
\begin{split}  
\tau_\mathrm{mag}^{-1}=M_\mathrm{res}\frac{\hbar^4\omega^{4}}{(\hbar^2\omega^{2}-\Delta^{2})^{2}}\frac{\mathrm{exp}(-\frac{\Delta}{k_{B}T})}{1+3\mathrm{exp}(-\frac{\Delta}{k_{B}T})}.
\label{eq:mag}
\end{split}
\end{equation}
In this framework, the parameter $M_\mathrm{res}$ characterizes the strength of the resonant scattering, while $\Delta$ denotes the energy gap of the magnetic excitations. The second term in Eq.~\ref{eq:mag} represents the scattering cross section at the resonant frequency $\Delta = \hbar \omega_\mathrm{res}$, and the third factor accounts for the thermal population of the magnetic excitations. By incorporating both the conventional phonon-scattering rate and the magnetic scattering rate according to $\tau(\omega,T)^{-1} = \tau_0^{-1} + \tau_\mathrm{mag}^{-1}$, the thermal conductivity in region II can be well reproduced within the Debye–Callaway model, as indicated by the pink solid line in Fig.~\ref{k-T}(b). From this analysis, an magnetic excitation gap of $\Delta = 5.2$ meV ($\sim60$ K) is extracted. Previous inelastic neutron-scattering measurements on \ce{Fe2SiO4} revealed four magnetic excitations at 3.3, 5.4, 5.9, and 11.4 meV in the antiferromagnetic state \cite{aronson2007}. These excitations exhibit weak dispersion, suggesting an origin in discrete energy levels arising from crystal-field effects and spin-orbit coupling \cite{aronson2007}. As illustrated in Fig.~\ref{k-T}(d), the cubic crystal field splits the energy levels of Fe$^{2+}$ ($3d^6$, $^5D_4$, 25-fold degenerate) ions into a low-lying triplet $t_{2g}$ and a doublet $e_g$, separated by a large energy gap of approximately 1.1 eV \cite{aronson2007}. The presence of noncubic symmetry at the Fe1 and Fe2 sites further lifts the degeneracy of the $t_{2g}$ levels, yielding excitation energies of $\sim91$ and $\sim186$ meV \cite{Burns1985,Ehrenberg_1993,aronson2007}. The remaining ground-state degeneracy is subsequently lifted by spin-orbit coupling over an energy scale of $\sim10$ meV \cite{Coey1981}. As a result, a total of eight excitations are expected within the spin-orbit manifold for the two crystallographically inequivalent Fe sites.  \ce{Fe2SiSe4} in region II and \ce{Fe2SiO4} in its antiferromagnetic state share closely related crystal and magnetic structures. Similar crystal-field- and spin-orbit-induced energy splittings are therefore expected in both materials, although further spectroscopic measurements are required to quantitatively determine the excitation spectrum of \ce{Fe2SiSe4}. Nevertheless, the energy gap $\Delta = 5.2$ meV extracted from our thermal-conductivity analysis closely matches the excitations at 5.4 and 5.9 meV in \ce{Fe2SiO4}, which have been assigned to the Fe2 and Fe1 sites, respectively \cite{aronson2007}. The corresponding resonant phonon frequency is approximately 1.3 THz, which lies in the typical range of acoustic phonons dominating heat transport.  Note that application of a magnetic field introduces additional Zeeman splitting, which can weaken or enhance resonant spin–phonon scattering and thereby suppress or enhance the thermal conductivity. Such behavior is commonly observed when the Zeeman energy becomes comparable to the magnetic excitation energy, as reported for \ce{K2V3O8} \cite{Sales2002}, \ce{CuGeO3} \cite{CuGeO1998}, \ce{SrCu2(BO3)2} \cite{Hofmann2001SrCu}, and $\alpha$-\ce{RuCl3} \cite{Hentrich2018}. In \ce{Fe2SiSe4}, however, the Zeeman energy induced by a magnetic field of 9 T ($\Delta_z = MB \sim 0.25$ meV) is an order of magnitude smaller than the energy scale associated with spin-orbit coupling. Here, a magnetization value of $M \sim 0.5$ $\upmu_\mathrm{B}$/Fe at 9 T around 60 K is used to estimate the Zeeman energy. Consequently, substantially higher magnetic fields would be required to induce significant changes in the thermal conductivity of \ce{Fe2SiSe4}.

\begin{table*}
\centering
\caption{Best-fit parameters of the Debye–Callaway model used to describe the thermal conductivity of \ce{Fe2SiSe4} in different temperature regions.}
\label{Debye-Callaway}
\renewcommand*{\arraystretch}{1.5}% set the height of each row to be the 1.5 text size 

\begin{tabularx}{\textwidth}{
>{\centering\arraybackslash}X
>{\centering\arraybackslash}X
>{\centering\arraybackslash}X
>{\centering\arraybackslash}X
>{\centering\arraybackslash}p{8em}
>{\centering\arraybackslash}X
>{\centering\arraybackslash}p{8em}
>{\centering\arraybackslash}X}
\hline
\hline
 & $v_s$ (m s$^{-1}$) & $d$ ($10^{-4}$ m) & $P_d$ $(10^{-42}$ $\mathrm{s}^{3})$ & $U_1$ $(10^{-17}$ $\mathrm{K}^{-1}$ $\mathrm{s})$ & $U_2$ & $M_\mathrm{res}$ $(10^{12}$ $\mathrm{s}^{-1}$) & $\Delta$ (meV) \\ 
\hline
$T<T_2$ & 4150.1(4) & 1.40(2) & 2.35(2) & 1.91(2) & 4.65(1) &  &\\
\hline
$T_2<T<T_1$ & 4150.1(4) & 1.40(2) & 11.78(3) & 0.69(1) & 1.01(2) & 1.01(1) & 5.2(1)\\
\hline
$T>T_1$ & 4150.1(4) & 1.40(2) & 31.05(5) & 0.11(2) & 5.49(2) &  &\\
\hline
\hline

\end{tabularx}
\end{table*}

\textit{Region III:}  Just below $T_2$, in the double-$\mathbf{q}$ magnetic state (region III), $\kappa_{xx}$ increases rapidly upon cooling and follows an approximate $T^{-1}$ temperature dependence, reaching a pronounced peak near 11 K. Such behavior is characteristic of phonon-dominated heat transport governed by conventional scattering mechanisms. Indeed, by considering only the conventional phonon-scattering rate $\tau_0^{-1}$, the temperature dependence of $\kappa_{xx}$ in region III can be well reproduced within the Debye–Callaway framework, as indicated by the red solid line in Fig.~\ref{k-T}(b). The peak at approximately 11 K corresponds to the crossover between the low-temperature boundary-scattering-dominated regime and the higher-temperature Umklapp-scattering-dominated regime. Evidently, the resonant spin–phonon scattering that dominates in region II is strongly suppressed in region III. As revealed by the thermal-expansion measurements, the lattice parameters undergo substantial changes at $T_2$ [see Figs.~\ref{Cp}(c,d)]. In addition, the magnetic structures in regions II and III differ markedly. The combined effects of abrupt structural and magnetic rearrangements significantly modify both the phonon spectrum and the energy-level splitting induced by crystal-field and spin-orbit interactions. The absence of phonons with energies resonant with the magnetic excitations in region III likely restores conventional phonon-scattering processes, thereby strongly enhancing the thermal conductivity.  Further evidence for the suppression of spin–phonon scattering in region III is provided by the steep increase in the phonon mean free path $\ell_\mathrm{ph}$ upon cooling just below $T_2$ [see the inset of Fig.~\ref{k-T}(c)]. The phonon mean free path is estimated using $\kappa_\mathrm{ph} = \frac{1}{3} C_{\mathrm{ph}} v_{\mathrm{s}} \ell_{\mathrm{ph}}$, where $v_{\mathrm{s}}$ is the average sound velocity, as commonly adopted for three-dimensional phonon-gas models under the assumption of a frequency-independent mean free path. The sound velocity is estimated via $v_s = \Theta_{\mathrm{D}} (k_B/\hbar) (6\pi^2 N / V)^{-1/3}$, with $N$ and $V$ denoting the number of atoms per unit cell and the volume of an unit cell, respectively. Although the Fe atoms in \ce{Fe2SiSe4} form quasi-one-dimensional sawtooth chains, a three-dimensional network is established through edge- and corner-sharing \ce{FeSe6} octahedra, justifying the use of a three-dimensional approximation. 

Interestingly, the phonon mean free path in region II is larger than that in region I, despite the presence of resonant spin–phonon scattering in region II. As shown in Figs.~\ref{VSM}(d,f) and \ref{Cp}(b), the magnetization for $B \parallel a$ and $B \parallel c$ deviates markedly from Curie–Weiss behavior below $\sim200$ K, and the magnetic specific heat extends well into the paramagnetic state. Both observations provide compelling evidence for the presence of pronounced short-range spin fluctuations in region I. Such fluctuations can effectively act as point defects and significantly reduce the phonon mean free path. Consistently, as summarized in Table~\ref{Debye-Callaway}, the fitted parameter associated with point-defect scattering, $P_d$, is substantially larger in region I than in regions II and III.  Strong suppression of thermal conductivity by short-range charge or spin fluctuations is not uncommon. For example, in various charge-ordered materials \cite{Smontara_CDW,Kuo2003_CDW,Murata2015_CDW,Gumeniuk2015_CDW,YangCDW,Yan2003_stripe,Hess1999_stripe} and low-dimensional or frustrated magnets \cite{Li_glass,Sharam_glass,Uehara2022_glass,Casto_glass,yang2023spin}, strong phonon scattering by short-range charge or spin fluctuations can even lead to glasslike thermal transport above the ordering temperatures. Upon entering the long-range ordered state, the suppression of short-range fluctuations leads to a rapid enhancement of thermal conductivity. These examples suggest that short-range fluctuations, even without static long-range order, can be very efficient phonon scatterers. Finally, we note that, compared to the broad maximum of approximately 4.8 W m$^{-1}$ K$^{-1}$ near 60 K, the low-temperature peak of about 23 W m$^{-1}$ K$^{-1}$ at 11 K is enhanced by a factor of $\sim5$. These results demonstrate that spin–phonon interactions provide an effective route for tuning phonon thermal conductivity, highlighting frustrated magnets as promising platforms for thermoelectric applications. Previous theoretical studies have suggested that \ce{Fe2GeS4} and \ce{Fe2GeSe4} are potential thermoelectric candidates \cite{FeGeCh2016}. However, their thermoelectric performance is typically limited by low electrical conductivity.

\section{Conclusions}

In conclusion, our thermal-expansion and thermal-transport measurements reveal strong spin–lattice coupling in the geometrically frustrated sawtooth-chain magnet \ce{Fe2SiSe4}, which exhibits a peculiar double-peak temperature dependence of the thermal conductivity. In the single-$\mathbf{q}$ antiferromagnetic state between $T_1 = 110$ K and $T_2 = 50$ K, resonant scattering of phonons by magnetic excitations originating from the spin-orbit manifold gives rise to a broad maximum in the thermal conductivity. Upon entering the double-$\mathbf{q}$ magnetic state below $T_2$, abrupt changes in the magnetic structure and the lattice parameters suppress the resonant spin–phonon scattering, resulting in a strong enhancement of the thermal conductivity and the emergence of a pronounced low-temperature peak. These findings establish \ce{Fe2SiSe4} as a compelling platform for exploring the intricate interplay among lattice, spin, and orbital degrees of freedom and demonstrate that spin–lattice coupling provides an effective route for tailoring thermal-transport properties in geometrically frustrated quantum magnets.

\medskip

\section*{Acknowledgements}
We thank Guiwen Wang and Yan Liu at the Analytical and Testing Center of Chongqing University for technical support. This work has been supported by National Natural Science Foundation of China (Grants Nos. 12474141, No. 12474148), Natural Science Foundation of Chongqing, China (Grant No. CSTB2025NSCQ-GPX0729), Fundamental Research Funds for the Central Universities, China (2025CDJ-IAISYB-034), the Venture and Innovation Support Program for Chongqing Overseas Returnees (Grant No. cx2024007), and Chinesisch-Deutsche Mobilitätsprogamm of Chinesisch-Deutsche Zentrum für Wissenschaftsförderung (Grant No. M-0496). Y.C. acknowledges the support by the National Natural Science Foundation of China (Grant Nos. 12374081, 12227806), the Open Research Fund of the Pulsed High Magnetic Field Facility (Grant No. WHMFC2024007), and Huazhong University of Science and Technology.

%\nocite{*}

\FloatBarrier

\end{document}